\documentclass[sigconf]{acmart}

\renewcommand\footnotetextcopyrightpermission[1]{} 

\AtBeginDocument{%
  }

\usepackage{array}
\usepackage{tabularx}
\newcolumntype{C}[1]{>{\centering\arraybackslash}p{#1}}
\usepackage{booktabs}
\usepackage[table]{xcolor}   
\usepackage{array}
\usepackage{makecell}        
\usepackage{caption}

\definecolor{good}{HTML}{CFEFCC}   
\definecolor{bad}{HTML}{F6C7C7}    
\definecolor{mid}{HTML}{F3E9B5}    

\usepackage{listings}
\usepackage{xcolor}

\usepackage[framemethod=tikz]{mdframed}
\newmdenv[
  topline=false,
  bottomline=false,
  rightline=false,
  linewidth=2pt,
  linecolor=gray!50,
  innerleftmargin=1.8pt,
  innerrightmargin=0pt,
  innertopmargin=0pt,
  innerbottommargin=0pt,
]{leftlinebox}

\usepackage{enumitem}
\usepackage{subcaption}
\usepackage{cleveref}
\definecolor{firebrick}{HTML}{B22222} 
\definecolor{clightgray}{rgb}{0.83, 0.83, 0.83}

\newcounter{mycodeboxcounter}

\usepackage[most]{tcolorbox}    	
\newtcolorbox{mycodebox}[1][]{
    standard jigsaw,
    opacityback=0,  
    boxrule = 0.pt,
    left=0pt,
    right=0pt,
    top=0pt,
    bottom=0pt,
    sharp corners, 
    width=\linewidth,      
    halign=flush left,     
    before skip=5pt,       
    after skip=5pt,        
    grow to right by=0mm,   
}
\usepackage{minted}
\usepackage{fancyvrb}

\newcommand{\vtextbf}[1]{\textbf{{\texttt{#1}}}}

\usepackage{xcolor}
\usepackage{listings}
\definecolor{codegray}{rgb}{0.5,0.5,0.5}
\definecolor{codeblue}{rgb}{0,0,1}
\definecolor{codegreen}{rgb}{0,0.5,0}
\definecolor{codepurple}{rgb}{0.58,0,0.82}
\definecolor{codered}{rgb}{0.7, 0.1, 0.2}
\definecolor{lightblue}{RGB}{230,240,255}
\lstdefinestyle{mystyle}{
    commentstyle=\color{codegray},
    keywordstyle=\bfseries\color{black},
    numberstyle=\footnotesize\color{black},
    stringstyle=\color{codepurple},
    basicstyle=\ttfamily\footnotesize,
    breakatwhitespace=false,         
    breaklines=true,                 
    captionpos=b,                    
    keepspaces=true,                 
    numbers=left,                    
    numbersep=5pt,                  
    showspaces=false,                
    showstringspaces=false,
    showtabs=true,                  
    tabsize=1,
    frame=single,
    framerule=0pt,
    rulecolor=\color{gray},
    xleftmargin=0pt,
    xrightmargin=0pt
  }
\usepackage{caption}
\begin{document}

\title{
Virtual-Memory Assisted Buffer Management\\In Tiered Memory
}

\author{Yeasir Rayhan and Walid G. Aref}
\affiliation{
  \institution{Purdue University, West Lafayette, IN, USA}
  \city{}
  \country{}
}
\email{{yrayhan, aref}@purdue.edu}

\renewcommand{\shortauthors}{}
\newcommand{\psys}{\textit{vmcache}}
\newcommand{\sys}{\textit{vmcache$^n$}}
\newcommand{\psyscall}{\texttt{move\_pages}}
\newcommand{\syscall}{\texttt{move\_pages2}}
\begin{abstract}
Tiered memory architectures have gained significant traction in the database community in recent years. In these architectures, the on-chip DRAM of the host processor is typically referred to as \textit{local memory}, and forms the primary tier. Additional byte-addressable, cache-coherent memory resources, collectively referred to as \textit{remote memory} (RMem, for short), form one or more secondary tiers. RMem is slower than local DRAM but faster than disk, e.g., NUMA memory located on a remote socket, chiplet-attached memory, and memory attached via high-performance interconnect protocols, e.g., RDMA and CXL. In this paper, we discuss how traditional two-tier (DRAM-Disk) virtual-memory assisted Buffer Management techniques generalize to an $n$-tier setting (DRAM-RMem-Disk). We present \sys{}, an $n$-tier virtual-memory-assisted buffer pool that leverages the virtual memory subsystem and operating system calls to migrate pages across memory tiers. In this setup, page migration can become a bottleneck. To address this limitation, we introduce the \syscall{} system call that provides \sys{} with fine-grained control over the page migration process. Experiments show that \sys{} can achieve up to 4$\times$ higher query throughput over \psys{} for TPC-C workloads.  

\end{abstract}

\maketitle

\section{Introduction} 
Most DBMSs implement a two-Tier (DRAM-Disk) Buffer Pool, where database pages\footnote{Throughout the rest of this paper, the term page refers to a database page unless stated otherwise.} are cached into DRAM for efficient access. Traditionally, DBMSs implement a Hash Table to index cached pages by their page identifier, i.e., PID. A PID of a cached page maps to a \textit{virtual} memory address of the corresponding page in memory. This Hash Table-based design introduces an additional indirection as each page access requires a Hash Table Lookup. This can incur significant overhead even when the workload is fully DRAM-resident and large, as the workload becomes CPU-bound due to hash table bloat, pointer chasing, latch contention, and reduced cache locality~\cite{HarizopoulosAMS08,ZhouLYS25}. 
To mitigate the overhead of Hash Table-based indirection, several alternatives have been proposed, e.g., pointer swizzling~\cite{GraefeVKKTLV14, LeisHK18, NeumannF20}, and virtual-memory assisted buffer management~\cite{LeisA0L023,RiekenbrauckWLR24}, LIPAH~\cite{OtakiBEG25,OtakiCEG25}. In this paper, we focus on virtual-memory assisted Buffer Pools, i.e., \psys{}. We examine how this design generalizes beyond the traditional two-Tier setting and extends to an $n$-tier storage architecture, where the first n-1 tiers are memory-based, e.g., DRAM, Persistent Memory, CXL Memory, and the final tier is disk-based, e.g., 
NVMe SSD, HDD. 

Despite the steady decline in DRAM prices until 2020~\cite{mac_trends,dram_trends}, recent market dynamics have led to a renewed spike in DRAM costs by up to 90\%~\cite{dram-spike}. Hence, \textit{memory stranding}, i.e., underutilized memory across servers has become an increasingly critical issue. Recent research, e.g.,~\cite{HaoZYS24,SunYYKSHJALJ0AX23,AhnCLGKJRPMK22,0001LBC24,memorycentric25,pasha25,LiuHWBNJNL25,Guo024,cxl_switch25,AravindJ24,WeisgutRSR25}, advocates for memory disaggregation and multi-tiered memory architectures to improve memory utilization and to reduce cost per effective giga-byte (GB) of memory. While early work demonstrates the feasibility of tiered memory architectures in buffer pool designs~\cite{ZhouAPC21,RenenLK18,HaoZYS24,RiekenbrauckWLR24}, the design space of virtual-memory assisted buffer management in a general $n$-tier setting remains largely unexplored. Given its success in two-tier designs, a natural next step is to understand their behavior and design trade-offs in an $n$-tier setting.


In this paper, we present \sys{}, an $n$-tier virtual-memory assisted Buffer Pool built on top of \psys{}. \sys{} inherits the core design principle of \psys{}, and enforces the following invariant: \textit{The  virtual address associated with a page remains fixed throughout its lifetime}. Similar to \psys{}, \sys{} delegates PID translation to the OS Page Table, and retains control over page promotion and page eviction from disk to memory tiers through \textit{libaio} interface~\cite{libaio}. 
The inclusion of additional memory tiers imposes an additional constraint on \sys{}, thereby extending the prior invariant: \textit{the physical frame backing a database page in memory must support dynamic mapping across different memory-resident tiers over time while ensuring that the corresponding virtual address of the page remains fixed throughout its lifetime}. In the remainder of this paper, we focus our discussion to 3-tier virtual-memory assisted buffer pools, where Tier-0, Tier-1, and Tier-2 correspond to a DRAM, a remote memory, and an NVMe disk, respectively. The techniques presented naturally generalize to an $n$-tier setting, where the intermediate layers may consist of multiple remote memory tiers. 

\sys{} retains control over page promotion and page demotion across memory tiers, and leverages page migration system calls, e.g., \texttt{mbind} and \texttt{move\_pages} to migrate pages between memory tiers. These system calls preserve a page's virtual address while updating the physical frame that backs it. The syscalls further install the virtual-to-physical mapping in the OS's Page Table, and thus making the physical location of the page frame transparent to the buffer manager without requiring any additional Hash Table Lookup. On modern hardware with fast memory tiers and NVMe SSDs, the overheads of the OS's memory management unit (MMU) and page migration system calls can become a performance bottleneck. Hence, we implement batching to amortize the migration cost of multiple pages. To further improve performance, we build on top of the native kernel system call \texttt{move\_pages} and propose \texttt{move\_pages2}. The \texttt{move\_pages2} custom system call enables \sys{} to retain fine-grained control over the page migration process with the following two knobs: \texttt{migration\_mode} and \texttt{nr\_max\_batched\_migration}. The \texttt{migration\_mode} knob lets \sys{} choose the strictness of the migration policy. The \texttt{nr\_max\_batched\_migration} knob lets \sys{} choose the maximum number of pages that can be batched together for migration.

\section{Virtual-Memory Assisted Buffer Management in Tiered Memory}
\label{sec:vmcache+}
By design, in \psys{}~\cite{LeisA0L023}, the virtual address associated with a page remains fixed throughout its lifetime. A page may be evicted from DRAM and later re-cached at the same virtual memory address. This stability allows \psys{} to safely delegate PID translation to the OS Page Table. 
If the page is DRAM-resident, the OS Page Table resolves the access to the corresponding physical frame. Otherwise, the OS raises a page fault, and \psys{} loads the page from disk. 
This design eliminates the additional Hash Table lookup cost, as page accesses are resolved through direct virtual-memory translation. 
While this is conceptually straightforward in a two-tier setting, the situation becomes more nuanced in an $n$-tier setting, where multiple memory tiers are involved. 

\noindent\textbf{\sys{}: An $n$-tier Virtual-Memory Assisted Buffer Pool. }
\sys{} extends  \psys{'s} two-tier design to an $n$-tier setting. Unlike \psys{} that maintains a single DRAM cache, \sys{} maintains separate caches for each memory tier. The key distinction is that \sys{}’s $n$-tier design requires the physical frame backing a page to be dynamically remapped across multiple memory-resident tiers without changing its virtual address. Next, we outline the design principles of \sys{}. 
\begin{enumerate}[leftmargin=*,label=\textbf{\arabic*.}]
    \item The fundamental invariant of an $n$-tier virtual-memory assisted Buffer Pool is a stable virtual addressing along with \textit{dynamically changeable physical frame mappings across the memory tiers}.
    \item Page migration can only update the physical frame backing a PID while preserving the virtual memory address of the PID.
    \item At any point in time, a page can reside in exactly one memory tier. Pages are not replicated across tiers due to the use of a single, fixed virtual address per PID.
    \item Due to the invariant of stable virtual addressing, only memory tiers configured in system-RAM mode are compatible with virtual-memory assisted buffer management~\cite{AravindJ24,WeisgutRSR25}. 
\end{enumerate}
\sys{} is incompatible with alternative memory-tier configurations, e.g., the Device Direct Access (DAX) mode~\cite{intel-cxl-3}. In DAX mode, the memory tier is exposed as a character device, and the virtual memory mappings created from the memory tier are permanently backed by physical frames from the corresponding device. This prevents remapping the virtual address to a different physical frame, i.e., the fundamental invariant of \sys{}.
\begin{table}[t]
  \captionsetup{aboveskip=-0.3pt}
  \caption{Conceptual comparison of buffer management techniques in tiered memory.}
  \label{tab:comparison}
  \small 
  \centering
\begin{tabular}{C{1.25cm}C{0.95cm}C{0.95cm}C{1.1cm}C{1.0cm}C{1.0cm}}
  \hline
  & \textbf{mmap} & \textbf{tradi.} & \textbf{Ptr swiz.} & \textbf{Hyrise} & \textbf{\em{vmcache$^n$}} \\
  &  & \cite{ZhouAPC21,HaoZYS24} & \cite{RenenLK18} & \cite{RiekenbrauckWLR24} & Sec.~\ref{sec:vmcache+} \\
  \hline
  PID-transl. & page tbl. & hash tbl. & invasive & page tbl. & page tbl. \\
  tier-trk. &  page tbl. & hash tbl. & hash tbl. &  page tbl.
   & page tbl. \\
  control & OS & DBMS & DBMS &  DBMS & DBMS \\
  mig-unit. &  OS page &  cacheline & cacheline & OS page & OS page\\
  mig-gran. &  $n$ &  \textit{n} & \textit{n} & 1 & \textit{n}\\
  data-mov. & \texttt{autonuma} & \texttt{memcpy} & \texttt{memcpy} & \texttt{mbind} & \texttt{move\_pages2}$^*$\\
  vir addr. & same & different & different & same & same\\
  page dup. & \cellcolor{good}no & \cellcolor{mid}yes  & \cellcolor{bad}yes & \cellcolor{good}no & \cellcolor{good}no\\
  var. size & \cellcolor{good}easy & \cellcolor{bad}hard & \cellcolor{bad}hard & 
  \cellcolor{good}easy & \cellcolor{good}easy \\
  graphs & \cellcolor{good}yes & \cellcolor{good}yes & \cellcolor{bad}no &  \cellcolor{good}yes & \cellcolor{good}yes
  \\
  implem. & \cellcolor{mid}med & \cellcolor{bad}easy & \cellcolor{bad}hard & 
  \cellcolor{good}easy & \cellcolor{good}easy \\
  \hline
  \end{tabular}
  \par\raggedright $^*$\sys{} also supports \texttt{mbind} and \texttt{move\_pages} 
  \vspace{-15pt}
\end{table}


\noindent \textbf{\sys{} vs. Other Buffer Management Techniques. }
Table~\ref{tab:comparison} presents a conceptual comparison of alternative buffer pool designs with \sys{}. 
\sys{} is built on top of \psys{}. 
It inherits the core advantages of \psys{} including support for variable-sized pages, graph workloads, and robust performance for both in-memory and out-of-memory workloads. In an $n$-tier setting, several new design dimensions arise including tracking the physical location of pages (tier-trk), defining the migration unit (mig-unit), determining migration granularity (mig-gran), specifying the data movement interface (data-mov), and handling page duplication (page dup). In  traditional designs~\cite{ZhouAPC21,HaoZYS24} and in pointer-swizzling approaches~\cite{RenenLK18}, the hash table explicitly tracks the physical location of each page across memory tiers. These designs support data movement at cache-line granularity using \texttt{memcpy} leading to page duplication across tiers. These designs may assign different virtual addresses to the same logical page over its lifetime. Hyrise adopts a virtual-memory assisted buffer pool design that relies on \texttt{mbind} to migrate pages individually, which does not scale in modern hardware. \sys{} performs batched migration of $n$ pages and leverages \syscall{} to enable efficient  page migration.

\noindent\textbf{2.1. Page Table Manipulation. }
Following \psys{}, at startup, \sys{} reserves a virtual memory address space equaling the size of the $n$-th Tier, i.e., disk. This reservation establishes a stable virtual address for every database page throughout its lifetime. To cache a page from disk into a memory-resident tier, \sys{} uses the \texttt{pread} system call. However, in an $n$-tier setting, \sys{} must support more flexible page placement than the traditional two-tier case, where pages are cached exclusively into DRAM. Precisely, \sys{} must be able to cache pages from disk into any memory-resident tier, not only DRAM. In addition, \sys{} must support transparent migration of pages across memory-resident tiers while preserving the page’s virtual address. 

\noindent\textbf{Adding pages from disk to a target memory-tier cache.} Following ~\cite{RiekenbrauckWLR24}, \sys{} uses the \texttt{mbind} system call to add a page to a target memory-tier cache. For example, we want to cache Page P3 in Tier-1. \texttt{mbind} sets the physical memory allocation policy for Page P3 to \texttt{MPOL\_BIND}. \texttt{MPOL\_BIND} ensures that the physical frame backing P3 is allocated from the memory tier specified by the \texttt{tgt\_tier} bitmask, i.e., 0b0010 (Tier-1) in the example.  In addition, \sys{} sets the \texttt{MPOL\_MF\_MOVE} flag to ensure that, if the page already resides in system memory, it is migrated to the memory tier indicated by the target-tier bitmask. When \texttt{mbind} 
succeeds, \sys{} reads Page P3 from disk via \texttt{pread} to the corresponding physical frame mapped by the virtual memory address. Then, the OS installs the virtual-to-physical mapping for P3 in the OS Page Table. Any subsequent access to P3 is served from Tier-1, until P3 is evicted or is moved to a different memory tier.
\begin{leftlinebox}
\begin{lstlisting}[language=C,basicstyle=\ttfamily\small, breaklines=true,numbers=none,tab=1]
u64 offset = 3*pageSize; u64 tgt_tier = 0b0010;
int mode = MPOL_BIND; int flags = MPOL_MF_MOVE;
mbind(virtMem+offset, pageSize, mode, tgt_tier, 8*sizeof(tgt_tier), flags);
pread(fd, virtMem+offset, pageSize, offset);
\end{lstlisting}
\end{leftlinebox}

\noindent\textbf{Adding and removing pages across memory-tier caches.} In an $n$-tier hierarchy, pages may migrate between two memory tiers. For example, when the working set exceeds the capacity of the DRAM-tier, pages can be evicted from DRAM and can be offloaded to a RMem tier rather than to disk. Similarly, pages can be prefetched from a RMem tier back to the DRAM tier. Migrations between intermediate RMem tiers are also possible to reduce a page's access latency. \sys{} maintains the following invariant during migration, i.e., \textit{although the physical frame backing a page may change, its virtual memory address, i.e., PID, must remain stable}. This invariant precludes software-level copy-based page migration mechanisms, e.g., \texttt{memcpy}, as it changes the virtual address of a page. 

\sys{} supports two system calls, namely, \texttt{mbind} and \texttt{move\_pages} to migrate pages between memory tiers. Both system calls preserve \sys{'s} invariant. 
Once the migration completes, the syscalls update the corresponding page-table entry of the Page Table to reflect the updated physical frame. 
Although both \texttt{mbind} and \texttt{move\_pages} can migrate pages, they differ in their migration granularity. While \texttt{move\_pages} can migrate multiple pages, \texttt{mbind} can only migrate one page at a time. For example, assume that we want to migrate Page P3 and Page P6 to Tier-1 memory. \sys{} needs to invoke \texttt{mbind} twice (i.e., separately) to migrate those pages.

\begin{leftlinebox}
\begin{lstlisting}[language=C,basicstyle=\ttfamily\small, breaklines=true,numbers=none,tab=2]
u64 offset1=3*pageSize; u64 offset2=6*pageSize;
int mode = MPOL_BIND; int flags = MPOL_MF_MOVE;
u64 tgt_tier = 0b0010;
for offsets in [offset1, offset2]: 
    mbind(virtMem+offset, pageSize, mode, tgt
    _tier, 8*sizeof(tgt_tier), flags);
\end{lstlisting}
\end{leftlinebox}
In contrast, \sys{} invokes the \texttt{move\_pages} system call once to migrate multiple pages (See the example below). The first argument (=0) refers to the calling process, i.e., \sys{}. The second argument (=2) sets the number of pages to migrate, and defines the size of the subsequent \textit{three} array arguments. The \texttt{pages} array contains the virtual addresses of the pages to be migrated, i.e., \{P3, P6\}. The \texttt{tgt\_tier} array specifies the destination memory tier for each page. The \texttt{status} array reports the migration status of each page upon migration completion. Finally, the \texttt{MPOL\_MF\_MOVE} flag requests the migration of the specified pages’ physical frames to the target memory tiers.

\begin{leftlinebox}
\begin{lstlisting}[language=C,basicstyle=\ttfamily\small, breaklines=true,numbers=none,tab=2]
u64[] pages = {virtMem+3*pageSize, virtMem+6*pageSize}; int flag = MPOL_MF_MOVE;
int[] tgt_tiers={1, 1}; int[] status={-1, -1};
move_pages(0, 2, offset, tgt_tier, status, flag);
\end{lstlisting}
\end{leftlinebox}

\noindent \textbf{2.2. Page States \& Synchronization. }
Following \psys{}, \sys{} uses 64 bits to represent the state of a page. The most significant 8 bits represent 4 page states in \psys{}, i.e.,  Unlocked (0), LockedShared (1-252), Locked (253), Marked (254), and Evicted (255). To accommodate additional memory tiers in an $n$-tier setting, \sys{} encodes the location, i.e., memory tier of the physical frame  backing the page using $\left\lceil \log_2(n-1) \right\rceil$ bits. For example, in a 3-tier setting (DRAM-RMem-Disk), 1 bit encodes the memory tier, thereby requiring a total of 9 bits to encode the state of a page. This yields tier-specific page states, e.g., UnlockedDRAM, UnlockedRMem, etc. 
Compared to \psys{}, \sys{} reserves fewer bits, i.e., 56-$\log_2(n-1)$ to encode the version counters. 

\begin{figure}[b]
    \captionsetup{aboveskip=-2pt}
    \centering
    \includegraphics[width=0.85\columnwidth]{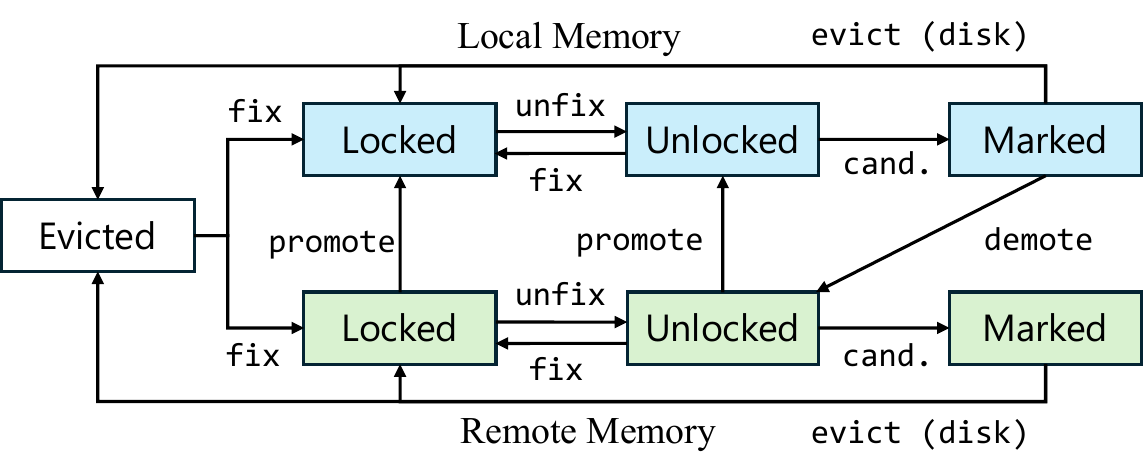}
    \caption{Page states in a 3-tier \sys{}.}
    \label{fig:page_state}
\end{figure}

\noindent \textbf{State transition. }
Figure~\ref{fig:page_state} shows the state transition diagram of a \sys{} page in a 3-tier setting. Unlike \psys{}, \sys{} maintains tier-specific page states to allow for flexible page movements across memory tiers. Following~\cite{ZhouAPC21}, \sys{} maintains \textit{four} migration flags to probabilistically migrate pages between memory tiers, i.e., \texttt{Dr, Dw, Rr, Rw} that we discuss below. Assume that \sys{} needs to access Page P1 that is in Evicted state. \sys{} calls \texttt{fix(P1)}. In a 3-tier setting, \texttt{fix(P1)} has \textit{two} possible target tiers: DRAM or RMem, chosen probabilistically by \texttt{Rr}. Assume that \sys{} chooses DRAM as the target tier. P1 transitions to the Locked State using a Compare-and-Swap operation (CAS, for short). \sys{} reads the page from disk and copies it to DRAM using the \texttt{mbind} system call. \texttt{mbind} implicitly updates the OS Page Table. After access, \sys{} calls \texttt{unfix(P1)}, and P1 transitions to the Unlocked State. When the DRAM cache utilization reaches a pre-defined threshold, e.g., 95\%, \sys{} pro-actively marks unlocked pages, e.g., P1 with the clock replacement algorithm. A Marked page in DRAM has two target tiers, i.e., RMem or disk, controlled by \texttt{Rw}. If \texttt{Rw} selects RMem, \sys{} uses the \texttt{move\_pages} system call to demote P1 from DRAM to the RMem. The syscall implicitly updates the OS Page Table with P1's updated resident memory tier. Page P1 starts fresh in State Unlocked in RMem. Otherwise, P1 is evicted to disk. In another scenario, assume that \sys{} needs to access Page P2 that resides in RMem. \sys{} invokes \texttt{fix(P2)}. Owing to the byte-addressability of RMem, \sys{} may either access P2 directly in place or first migrate it to DRAM before accessing it. This decision is made by \texttt{Rr}. Suppose that \texttt{Rr} chooses to promote P2 to DRAM. In this case, \sys{} prefetches a set of \texttt{Unlocked} pages from the remote tier and bundles them with P2, promoting all selected pages together in a batched migration.

\noindent \textbf{2.3. Page Replacement. }
\label{sec:page_replace}
\sys{} maintains a separate cache for each memory tier. When all the caches get full, \sys{} needs to evict pages from them. Following \psys{}, \sys{} uses the clock replacement algorithm for this purpose. Unlike \psys{}, in a 3-tier setting, there are \textit{two} distinct eviction paths, i.e., eviction between memory tiers  and eviction from a memory tier to disk. In addition, the introduction of additional memory tiers introduces additional promotion paths between two memory tiers. \sys{} uses a simple heuristic to move pages upwards in the 3-tier hierarchy. 

\noindent \textbf{Batch eviction between memory tiers. }
Unlike \psys{}, each cache in \sys{} maintains a separate resident set, i.e., an open-addressing hash table that indexes the resident page PIDs in the corresponding memory tier. For example, in a 3-Tier setting, when DRAM cache size reaches a threshold and the target tier is RMem, \sys{} starts the following eviction routine: (1)~Scan the DRAM resident set, collect the set of Marked pages selected for eviction, and lock these pages. (2)~Update the location of the pages in the OS Page Table using \texttt{move\_pages}. Since the pages remain memory-resident, no additional handling is required for dirty pages. (3)~Remove the locked pages from the DRAM resident set and insert them into the RMem resident set. (4)~Update the physical tier location metadata for each page and release the locks.

\noindent \textbf{Batch promotion between memory tiers. }
In \sys{}, a page may reside in memory, but in a RMem tier. 
When a page lookup requires accessing a page from the RMem tier, \sys{} probabilistically decides whether the page should be promoted. If promotion is triggered, \sys{} locks the target page, scans the corresponding resident set, and collects a set of \texttt{Unlocked} pages selected for promotion. \sys{} locks these pages, and follows Steps 2--4 of the batch eviction routine.

\section{\vtextbf{move\_pages2}: Efficient Page Migration Across Memory Tiers}
\sys{} relies on the virtual memory subsystem and OS-driven system calls, e.g., \texttt{mbind}~\cite{linux_mbind} and \texttt{move\_pages}~\cite{linux_move_pages}, to dynamically alter the physical location of a page while preserving its virtual address. In any virtual-memory assisted buffer pool operating in an $n$-tier setting, \textit{page migration} lies on the critical path of query execution. Hence, the performance of these system calls is paramount to achieving an efficient virtual-memory-assisted buffer pool. Although the native page migration system calls suffice in general, they may encounter performance and scalability issues under certain workloads. To address this limitation, we propose \syscall{}. 

\noindent\textbf{3.1. Motivation. }
Linux handles page migration through two system calls, \texttt{mbind}~\cite{linux_mbind} and \psyscall{}~\cite{linux_move_pages}. We focus on \psyscall{} as it enables batched page migration. In contrast, \texttt{mbind} migrate pages individually, and does not scale. 

\noindent \textbf{1. TLB Shootdowns. } 
\psyscall{} takes a list of pages as input and migrates them in multiple rounds. In each round, the kernel groups up to \verb|NR_MAX_BATCHED_MIGRATION| consecutive pages targeting the same memory tier, batches them, and enqueues them into a \verb|pagelist| queue. The batch size directly impacts performance. Larger batches reduce TLB invalidation overhead and decrease the number of invocations of the \verb|store_status| helper function that updates per-page status. TLB shootdowns are expensive because they require inter-processor interrupts (IPIs) to invalidate TLB entries across CPUs. By default, in \psyscall{}, \verb|NR_MAX_BATCHED_MIGRATION| is set to 512. The default migration mode is \verb|MIGRATE_SYNC|, which may trigger up to \verb|NR_MAX_BATCHED_MIGRATION| TLB shootdowns per round, significantly degrading migration performance.

\noindent \textbf{2. Abort-on-failure strategy. }
During migration, \psyscall{} invokes the kernel function \texttt{do\_pages\_move}, whose error-handling strategy has significant performance implications. When the kernel encounters a user-space access failure, an invalid target node, or a permission error for any page, it immediately aborts the operation. Before aborting, it migrates only the pages accumulated in the current batch, skipping all remaining pages in the input \texttt{pages} list, even if they are eligible for migration. This ``abort-on-failure" strategy is particularly detrimental in multi-threaded environments, where concurrent activity increases the likelihood of pages being locked, under writeback, or temporarily unavailable. 

\noindent\textbf{3.2. Design Principle \& Implementation Details. }

\noindent \textbf{\syscall{}.} \syscall{} excludes the \texttt{pid} and \texttt{flags} parameters from the original \psyscall{} interface and instead introduces two new parameters, \texttt{migrate\_mode} and \texttt{nr\_max\_batched\_migration}, to provide \sys{} with fine-grained control over page migration. By default, \syscall{} sets \texttt{pid = 0} and \texttt{flags = MPOL\_MF\_MOVE}, thereby granting \sys{} exclusive control over the migration process. The signature of the proposed custom \syscall{} implementation is as follows.

\begin{mycodebox}
\begin{small}
\raggedright
\begin{Verbatim}[commandchars=\\\{\}]
long {move_pages2} (unsigned long {count}, void {*pages[.count]}, 
    const int {nodes[.count]}, int {status[.count]}, 
    enum migrate_mode {mode}, int {nr_max_batched_migration});
\end{Verbatim}
\begin{itemize}
    \item \texttt{count}: The number of pages to process.
    \item \texttt{pages}: An array of page-pointers that need to be processed.
    \item \texttt{nodes}: An array of target NUMA node IDs.
    \item \texttt{status}: An array to store the migration status of each page.
    \item \texttt{mode}: The mode of migration. 
    \item \texttt{nr\_max\_batched\_migration}: The maximum number of pages that can be accumulated before TLB shootdown is invoked.
\end{itemize}
\end{small}
\end{mycodebox}

\noindent\textbf{1. Reducing TLB Shootdown.}
\syscall{} introduces \texttt{nr\_max\_batched\_migration} as a parameter to regulate the number of pages migrated in a single batch. This parameter directly impacts migration performance by amortizing the cost of TLB shootdowns up to \texttt{nr\_max\_batched\_migration} pages. In addition, \syscall{} exposes three page migration modes through the \texttt{migration\_mode} parameter.
\begin{itemize}
    \item \verb|MIGRATE_ASYNC| performs asynchronous migration, i.e., it proceeds with the next page even if the current page migration fails, making it a non-blocking approach.
    \item \verb|MIGRATE_SYNC| performs synchronous migration, i.e., 
    the kernel
    blocks until the current page migration succeeds.
    \item \verb|MIGRATE_SYNC_LIGHT| 
    does not block on page writebacks to reduce the stall time. 
\end{itemize}

\noindent\textbf{2. Optimistic Failure Handling.} 
\syscall{} adopts an optimistic error-handling strategy by allowing partial migration. Invoking a system call incurs the overhead of crossing the OS-\sys{} boundary. \syscall{} amortizes this overhead by migrating as many pages as possible within each invocation rather than terminating prematurely due to an error. 

\noindent\textbf{3.3. Implementation Details. }
We introduce approximately 150 lines of code 
changes across 
the following four kernel functions: \verb|do_pages_move|, \verb|move_pages_and_store_status|, \verb|do_move_pages_to_node|, and \verb|migrate_pages| to implement \syscall{}. We implement \syscall{} in Linux Kernel 6.8.0 on Ubuntu 24.04 LTS. The kernel disk image is available on CloudLab~\cite{DuplyakinRMWDES19} and can be accessed here\footnote{\href{}{urn:publicid:IDN+utah.cloudlab.us+image+pmoss-PG0:LINUX6.8.12\_MIGRATE}}. 

\noindent\textbf{1. Migration Modes and Bulk Batching. }
Our custom implementation of \verb|migrate_pages| (cf. Listing~\ref{lst:migrate_pages2}) enables \sys{} to override the default Linux migration mode, i.e., \verb|MIGRATE_SYNC|. For \verb|MIGRATE_SYNC|, the kernel follows the default blocking path (Line~\ref{mig:choose_mode1}). Otherwise, the kernel follows the non-blocking path (Line~\ref{mig:choose_mode2}). 
\syscall{} further allows \sys{} to dynamically select the maximum batch size during migration (Lines~\ref{mig:cut_1},~\ref{mig:cut_2}).
\begin{lstlisting}[
    caption={Pseudocode of \texttt{migrate\_pages2}. 
    % in \texttt{move\_pages2}. The modifications introduced over the original implementation are highlighted.
    }, 
    label={lst:migrate_pages2},
    mathescape=true
]
int migrate_pages2 ($\ldots$, @enum migrate_mode mode@, $\ldots$, @int nr_max_batched_migration@){
again:
  nr_pages = 0; // Iterate over the pages in "from"
  list_for_each_entry_safe(folio, folio2, from, lru) { 
    nr_pages += folio_nr_pages(folio);
    if (nr_pages >= @nr_max_batched_migration@) break;(*|\label{mig:cut_1}|*)
  }
  // Batch the pages in "from" into folios and migrate
  if (nr_pages >= @nr_max_batched_migration@)(*|\label{mig:cut_2}|*)
    list_cut_before(&folios, from, &folio2->lru);
  else
    list_splice_init(from, &folios);
  @if (migration_mode == MIGRATE_SYNC)(*|\label{mig:choose_mode1}|*)
    rc = migrate_pages_sync($\ldots$, MIGRATE_SYNC, $\ldots$); 
  else (*|\label{mig:choose_mode2}|*)
    rc = migrate_pages_batch($\ldots$, mode, $\ldots$, 
          nr_max_batched_migration);@
  // Store non-migrated pages in "ret_folios" for retry
  list_splice_tail_init(&folios, &ret_folios);  
  if (rc < 0) {
    rc_gather = rc; list_splice_tail(&split_folios, &ret_folios); goto out;
  }
  rc_gather += rc;
  if (!list_empty(from)) goto again; // Prep next folios
 out:
  // Move back the non-migrated pages in the current 
  // migration round to "folios" for potential 
  // retry in the next migration round
  list_splice(&ret_folios, from); 
  if (list_empty(from)) rc_gather = 0;
  return rc_gather;
}
\end{lstlisting}
\noindent\textbf{2. Optimistic Error Handling. }
Our custom implementation of Kernel Function \texttt{do\_pages\_move} (cf. Listing~\ref{lst:cus_do_pages_move}) ensures that if the kernel encounters a user-space access failure, an invalid target node, or a permission issue for a particular page, it does not abort immediately. Instead, it records the error in the \texttt{status} array and invokes the custom Kernel Function \verb|move_pages_store_status2| to migrate the pages accumulated in the current round. After completing the batch, the kernel initiates a new migration round and continues processing the remaining pages in the \texttt{pages} list (Lines~\ref{new_sys:handle_error_start}--\ref{new_sys:handle_error_end}). \syscall{} follows the same routine when it encounters the errors during migration (Lines~\ref{new_sys:migrate_error_start}--\ref{new_sys:migrate_error_end}).

\begin{lstlisting}[
    style=mystyle, 
    caption={Pseudocode of \texttt{do\_pages\_move2}.
    % in \texttt{move\_pages2}. The modifications introduced over the original implementation are highlighted.
    },  
    label={lst:cus_do_pages_move},
    tabsize=1,
    basicstyle=\ttfamily\footnotesize,
    mathescape=true
]
static int do_pages_move2 ($\ldots$,@enum migrate_mode mode,int nr_max_batched_migration@){
  for (i=start=0; i<nr_pages; i++) { (*|\label{sys:loop_start}|*)
    @if (error in copying pages[i] or nodes[i])
        goto handle_error;
    if (error in handling target tier) 
        goto handle_error;@
    if (current_node == NUMA_NO_NODE) { // Start mig.
      current_node = node; start = i;
    } else if (node != current_node) { (*|\label{sys:mig_end2}|*)
      // End the current migration round and migrate 
      // the pages collected in the current round
      @err = move_pages_and_store_status2($\ldots$, mode, nr_max_batched_migration);@ 
      @if (err) goto migrate_error;@
      start = i; current_node = node;
    }
    // Queue page p for the current migration round 
    err = add_page_for_migration();(*|\label{sys:mig_queue}|*)if (err > 0) continue;(*|\label{sys:mig_end1}|*) 
    err = store_status(); 
    @migrate_error:(*|\label{new_sys:migrate_error_start}|*)
      if (i + 1 == nr_pages || err < 0) { 
        err1 = move_pages_and_store_status2(); // Migrate
        if (err >= 0) err = err1;
        current_node = NUMA_NO_NODE; // New mig. round
      }
      continue; // Continue migration (*|\label{new_sys:migrate_error_end}|*)
    handle_error:(*|\label{new_sys:handle_error_start}|*)
      err1 = store_status(status, i, err, 1);
      if (err1) err = err1;
      err1 = move_pages_and_store_status2(); //Migrate
      if (err >= 0) err = err1;   
      current_node = NUMA_NO_NODE; // New round@ (*|\label{new_sys:handle_error_end}|*)
  }
out_flush:
  @if (current_node != NUMA_NO_NODE) {
    err1 = move_pages_and_store_status2($\ldots$, mode, nr_max_batched_migration);
    if (err >= 0) 
        err = err1;@
  }
out: 
  return err;
}
\end{lstlisting}


\section{Performance Evaluation}
All experiments run on a CloudLab~\cite{DuplyakinRMWDES19} sm220u node, i.e., an Intel(R) Xeon(R) Silver 4314 CPU with 2 NUMA sockets, each with 32 logical threads. 
We treat DRAM of NUMA Socket 0 as local memory (Tier-1), DRAM of NUMA Socket 1 as remote memory (Tier-2) and a 960 GB Samsung PCIe4 NVMe as disk (Tier-3). All experiments use 32 threads of NUMA Socket 0. Following~\cite{LeisA0L023},  we use the same TPC-C and a key-value workload that consists of random point lookups. The key–value workload uses 8-byte uniformly distributed keys and 120-byte values. The TPC-C and random-read workloads span approximately 190 GB and 130 GB, respectively. 

\begin{figure*}
    \captionsetup[subfigure]{aboveskip=-2pt}
    \captionsetup[subfigure]{belowskip=-2pt}
    \begin{subfigure}[t]{0.70\columnwidth}
        \centering
        \includegraphics[width=\columnwidth]{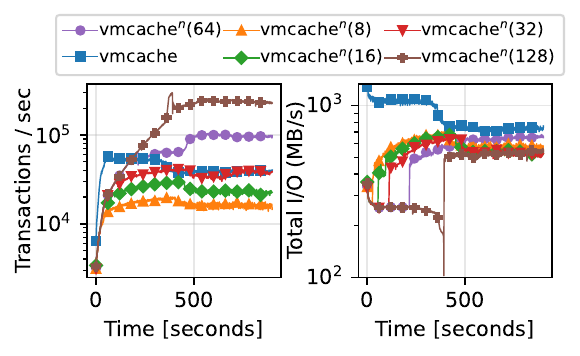}
        \caption{\sys{} vs. \psys{} (TPC-C)}
        \label{fig:disk_vs_remote_tpcc}
    \end{subfigure}
    \begin{subfigure}[t]{0.70\columnwidth}
        \centering
        \includegraphics[width=\columnwidth]{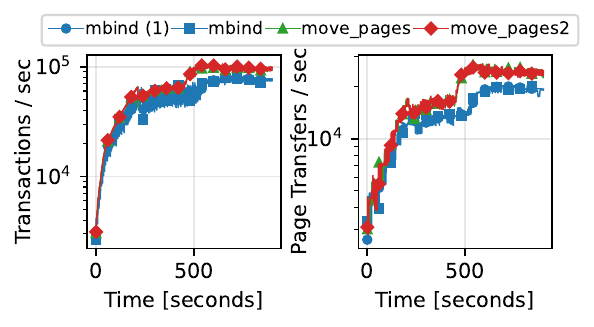}
        \caption{\syscall{} performance (TPC-C).}
        \label{fig:mp2_vs_mig_tpcc}
    \end{subfigure}
    \begin{subfigure}[t]{0.35\columnwidth}
        \centering
        \includegraphics[width=\columnwidth]{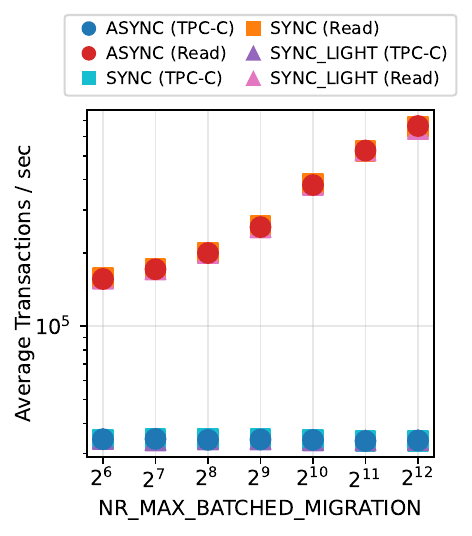}
        \caption{Max migrations.}
        \label{fig:abl_max_batch}
    \end{subfigure}
    \begin{subfigure}[t]{0.70\columnwidth}
            \centering
        \centering 
         \includegraphics[width=\columnwidth]{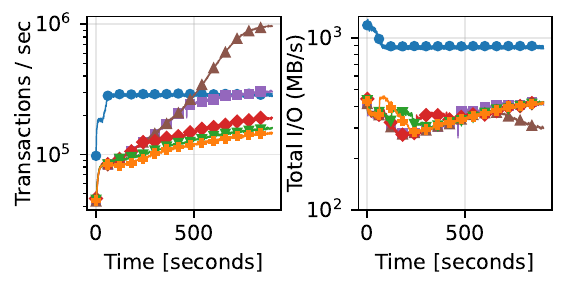}
    \caption{\sys{} vs. \psys{} (Random read)}
    \label{fig:disk_vs_remote_rndread}
    \end{subfigure}
    \begin{subfigure}[t]{0.70\columnwidth}
        \centering 
         \includegraphics[width=\columnwidth]{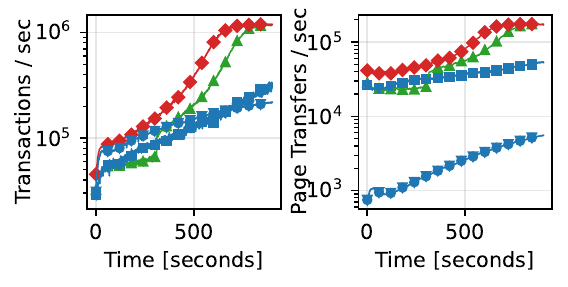}
    \caption{\syscall{} performance (Random read)}
    \label{fig:mp2_vs_mig_rnd}
    \end{subfigure}
    \begin{subfigure}[t]{0.35\columnwidth}
        \centering 
         \includegraphics[width=\columnwidth]{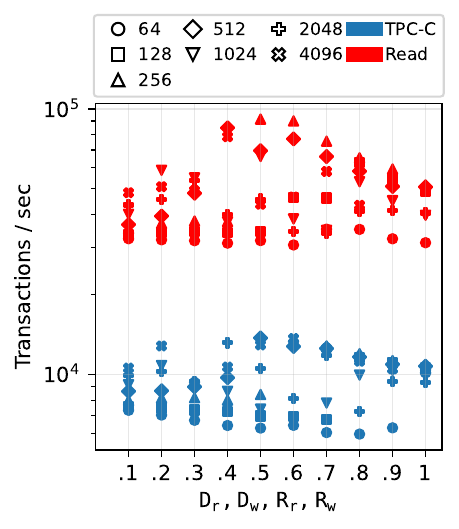}
    \caption{Migration ratio.}
    \label{fig:abl_mig-flag}
    \end{subfigure}
\captionsetup{
belowskip=-10pt, 
}
\caption{\sys{} vs. \psys{}.}
\label{fig:exp}
\end{figure*}

\noindent\textbf{\sys{} vs. \psys{}. }
~\Cref{fig:disk_vs_remote_tpcc,fig:disk_vs_remote_rndread} compare the performance of \sys{} and \psys{} over time for the TPC-C and random-read workloads. The local memory capacity is fixed at 32 GB, while the remote memory capacity varies from 8 GB to 128 GB. Initially, at time $T_0$, the remote memory tier is empty. As transactions execute over time, it gradually becomes populated. For TPC-C, \sys{} achieves up to 
1.67$\times$ and 3.82$\times$ higher throughput than \psys{} when the remote memory capacity is 2$\times$ and 4$\times$ the size of local memory, respectively. Although \sys{} reduces disk I/O, when the remote memory is small, traffic between memory tiers becomes the performance bottleneck. For the random-read workload, with no writes, the performance gain is not significant, e.g., \sys{} achieves 1.36$\times$ better transactions throughput when remote memory is 4$\times$ the size of the local memory. Here, page transfers between memory tiers dominate the execution cost as the working set exhibits random access patterns. 

\begin{figure}[b]
    \centering
    \begin{subfigure}[t]{0.45\columnwidth}
        \centering
        \includegraphics[width=\columnwidth]{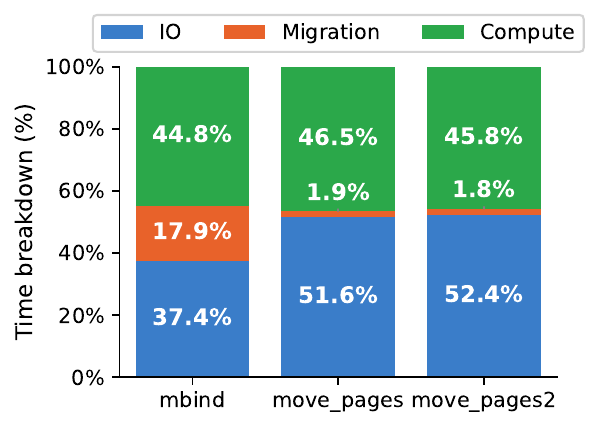}
        \caption{TPC-C workload.}
        \label{fig:brk_tpc-c}
    \end{subfigure}
    \begin{subfigure}[t]{0.45\columnwidth}
            \centering
        \centering 
         \includegraphics[width=\columnwidth]{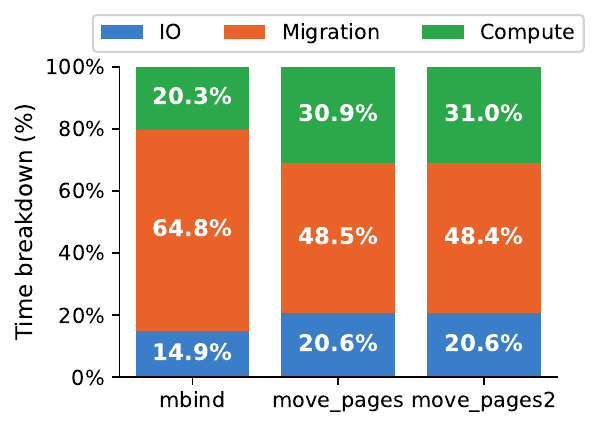}
    \caption{Random read workload.}
    \label{fig:brk_rnd-read}
    \end{subfigure}
\captionsetup{belowskip=-10pt, aboveskip=-0.1pt}
\caption{Execution time breakdown of \sys{}.}
\label{fig:brk}
\end{figure}

\noindent\textbf{Where does the time go? }
Figure~\ref{fig:brk} gives the execution time breakdown of \sys{} for the TPC-C workload (Fig.~\ref{fig:brk_tpc-c}) and the random-read workload (Fig.~\ref{fig:brk_rnd-read}) using alternate page migration syscalls. Disk I/O dominates the TPC-C workload, whereas page transfers between memory tiers dominate the random-read workload. Using \texttt{mbind} incurs substantial overhead for the random-read workload, accounting for 64.8\% of the total execution time, as \texttt{mbind} does not support bulk page movement. Even \psyscall{} and \syscall{} incur significant overheads of 48.5\% and 48.4\%, respectively, for the random-read workload. These results highlight the critical importance of an efficient page-movement interface for virtual-memory assisted buffer pools in an $n$-tier memory hierarchy. For these experiments, the local and remote memory capacity is set to 32 and 64 GB, respectively. The migration ratios and the batch eviction size between memory tiers are set to 1 and 512, respectively.

\noindent\textbf{Performance evaluation of \syscall{}.} 
~\Cref{fig:mp2_vs_mig_tpcc,fig:mp2_vs_mig_rnd} give the performance of \syscall{} in \sys{} against baseline page migration system calls, i.e., \texttt{mbind} and \texttt{move\_pages}. \texttt{mbind(1)} sets the batch eviction size between memory tiers to 1. For the rest, the batch eviction size is set to 512. \syscall{} exhibits similar performance to \psyscall{} for the TPC-C workload (cf.~\ref{fig:mp2_vs_mig_tpcc}). However, for the random-read workload, \syscall{} achieves 1.42$\times$, 1.32$\times$ better query and page migration throughputs over \psyscall{}, respectively (cf.~\ref{fig:mp2_vs_mig_rnd}), due to the maximum number of memory pages \syscall{} can batch together in a single round compared to the baselines, thus amortizing the cost of TLB invalidation. 

\noindent\textbf{Ablation analysis of \syscall{}.} 
Figure~\ref{fig:abl_max_batch} gives the performance of \syscall{} under varying \texttt{|NR\_MAX\_BATCHED\_MIGRATION|} values and migration modes. \texttt{|NR\_MAX\_BATCHED\_MIGRATION|} decides the maximum number of memory pages that can be migrated in a single round. In \sys{}, this is set to 2$\times$ the batch eviction size between memory tiers. For the random-read workload, increasing this value improves performance, as larger batches amortize migration overhead. For TPC-C, which is write-dominated, increasing the batch size does not provide similar benefits. Migration modes \texttt{ASYNC}, \texttt{SYNC} and \texttt{SYNC\_NO\_COPY} yield similar performance benefit in \sys{}, as the pages are locked before migration, regardless. Hence, allowing partial page migration does not translate into measurable performance improvement in \sys{}. Figure~\ref{fig:abl_mig-flag} gives the performance of \sys{} under different migration ratios. While a larger \texttt{|NR\_MAX\_BATCHED\_MIGRATION|} improves performance when page migrations are infrequent, it does not provide the same benefit as the migration frequency increases.

\section{Concluding Remarks}
We present \sys{}, an $n$-tier virtual-memory assisted buffer pool that delivers up to 4$\times$ higher TPC-C throughput than \psys{}.
The key findings can be summarized as follows.

\noindent 1. For virtual-memory assisted buffer pools, remote memory investment has a clear cost break-even point between, when the remote memory capacity is between 1$\times$ and 2$\times$ DRAM capacity. Below this threshold, the net gain in QPS/\$ compared to a 2-tier buffer pool is negative, as page migration overheads between memory tiers dominate any capacity benefit. Beyond the break-even point, the net gain in QPS/\$ becomes strongly positive and increases with scale, reaching +41 QPS/\$ and +87 QPS/\$
\footnote{Assuming, 
a Micron DDR4 3200MHz ECC RDIMM Memory is priced at 175\$
}, 
when remote memory capacity is 2$\times$, 4$\times$ DRAM capacity, respectively. 

\noindent2. Page migration between memory tiers is the bottleneck in an $n$-tier virtual-memory assisted buffer pool, particularly when the working set mostly fits within the memory tiers. \syscall{} mitigates this overhead by amortizing the migration cost through large batched transfers per each migration round. However, page migration overheads still remain. In our experiments, less than 0.005\% of execution time is attributed to kernel–user mode transitions and migration preparation overheads~\cite{XingB25}. Thus, re-architecting the system around microkernels to manage pages and page tables in user space is unlikely to address the dominant bottleneck~\cite{LeisD24}. Rather, more efficient page migration implementations are likely to yield greater benefit, be it in user space~\cite{SchuhknechtR25} or kernel space~\cite{YanLNB19, XiangLD0RY024,AwadBBSL17,Amit17,GandhiKACHMNSU16,CoxB17}.

\noindent3. Due to the invariant of \textit{stable virtual addresses with alternating physical memory tiers}, an $n$-tier virtual-memory assisted buffer pool can include remote memory tiers only if they are exposed to the host server in System-DRAM mode. Memory allocated via DAX or App Direct mode cannot be used in an $n$‑tier virtual-memory assisted buffer pool,  
which limits the applicability of virtual-memory assisted buffer pools in this setting.

\bibliographystyle{ACM-Reference-Format}
\bibliography{sample-base}

\end{document}